\documentclass[osajnl,twocolumn,showpacs,superscriptaddress,10pt]{revtex4-1} %% use 11pt for Applied Optics
\usepackage{amsmath,amssymb,graphicx}

\def\F#1{\mathcal{F}\left\{ {#1} \right\}}

\begin{document}
\title{Spatiotemporal characterization of ultrashort optical vortex pulses}

\author{Miguel~Miranda}\email{Corresponding author: miguel.miranda@fysik.lth.se}
\author{Marija~Kotur}
\author{Piotr~Rudawski}
\author{Chen~Guo}
\author{Anne~Harth}
\author{Anne~L'Huillier}
\author{Cord~L.~Arnold}

\affiliation{Department of Physics, Lund University, P.O. Box 118, SE-221 00 Lund, Sweden}


\begin{abstract}
Generation of few-cycle optical vortex pulses is challenging due to the large spectral bandwidths, as most vortex generation techniques are designed for monochromatic light.
In this work, we use a spiral phase plate to generate few-cycle optical vortices  from an ultrafast titanium:sapphire oscillator, and characterize them in the spatiotemporal domain using a recently introduced technique based on spatially resolved Fourier transform spectrometry.
The performance of this simple approach to the generation of optical vortices is analyzed from a wavelength dependent perspective, as well as in the spatiotemporal domain, allowing us to completely characterize ultrashort vortex pulses in space, frequency, and time.
\end{abstract}

\maketitle

\section{Introduction}

An optical vortex is a field with a phase singularity, such that the phase is undefined in the center of the vortex and thus the intensity must be zero. In addition, the phase variation along a path enclosing the singularity is non-zero:  
\begin{equation}
\ell=\frac{1}{2\pi}\oint \! \nabla \varphi(\mathbf{r})\cdot d\mathbf{r},
\end{equation}
where \(\ell\), which must be an integer different from $0$, is called the topological charge.
In the simplest case, the phase front of the light twists helically like a corkscrew around the singularity, according to \(\varphi(\rho,\theta)=\ell\theta\),  where \(\rho\) and \(\theta\) are  the radial and angular components in cylindrical coordinates.

Phase singularities in propagating wave fields were for the first time observed by Nye and Berry in ultrasound fields in the 1970s \cite{NyePotRSoLAMPaES1974}. Vortices in laser beams were later described \cite{CoulletOC1989} and it was recognized by Allen and co-workers in 1992 that helically phased beams carry orbital angular momentum (OAM) \cite{AllenPRA1992,Allen1999}. The particular properties of optical vortex beams have led to many applications \cite{YaoAOP2011}, such as imaging of faint objects around a bright one \cite{SwartzlanderOL2001}, overcoming the diffraction limit in stimulated emission depletion microscopy \cite{BretschneiderPRL2007}, and optical tweezers \cite{ONeilOC2000}.

Several methods exist today to convert a Gaussian laser beam into a vortex beam.
The most commonly used are based on phase masks which are simply inserted into the beam in transmission \cite{BeijersbergenOC1994}, diffraction holograms \cite{BazhenovJoMO1992}, and refractive–diffractive elements \cite{BockOL2012,MusigmannAO2014}.  Recent advances on spatial-light modulators enable the production of optical vortices with added flexibility, such as easily changing target wavelength and topological charge.

\begin{figure*}[htbp]
\centering
%\fbox{\includegraphics[]{schematic.pdf}}
\includegraphics[]{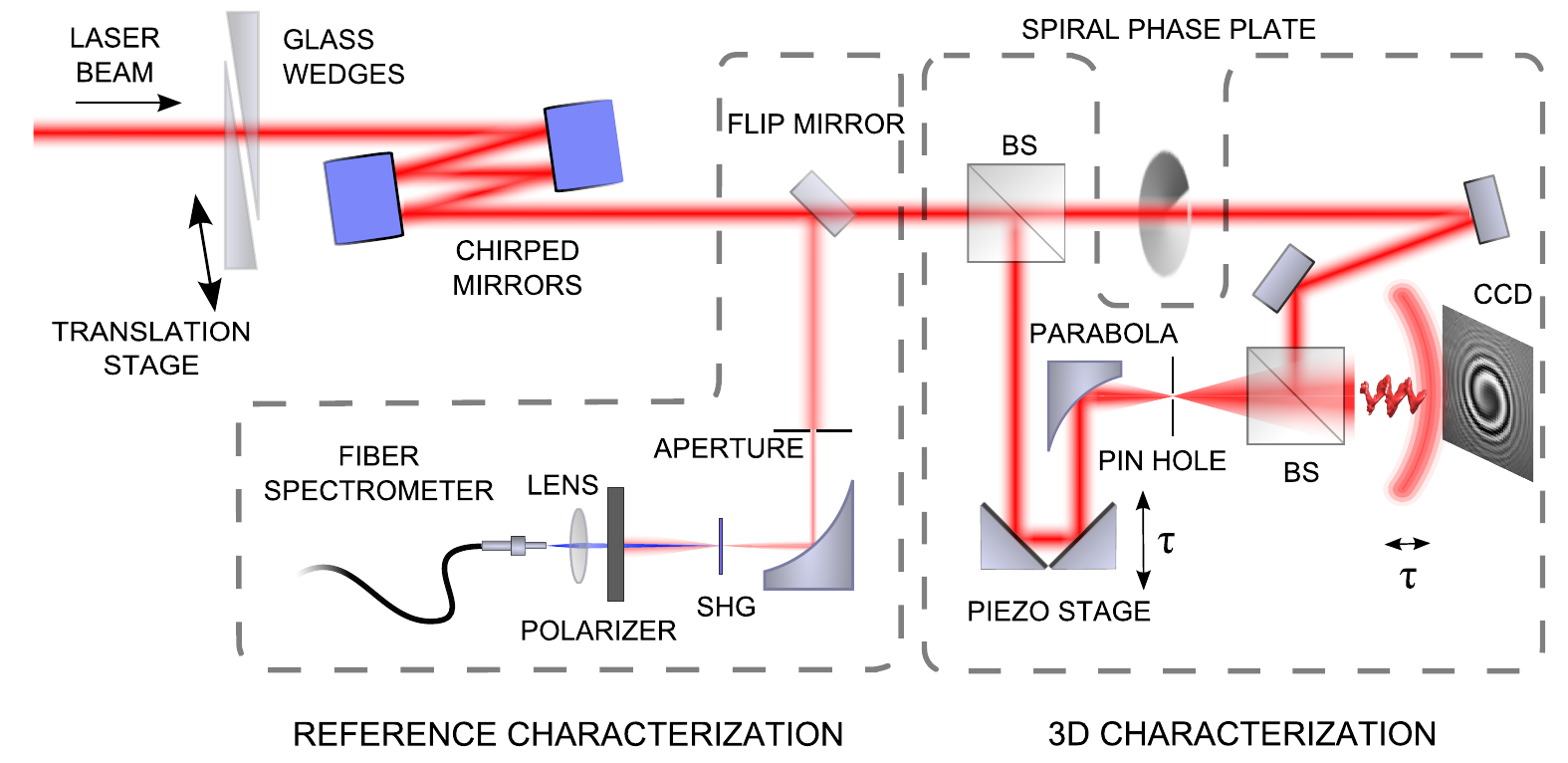}
\caption{Simplified experimental setup. The beam from an ultrafast oscillator is compressed with a chirped mirror and fused silica wedge compressor. The tunable dispersion of the compressor is also used for the temporal characterization of the reference pulse with the d-scan technique. A dispersion-balanced interferometer, based on a Mach–Zehnder design, is used to interfere a spatially-filtered reference wave with the generated vortex pulse on the chip of a CCD camera. The delay between the two fields is scanned and an interferogram is recorded for each pixel.}
\label{fig:setup}
\end{figure*}

Recently, optical vortices  have attracted interest in the ultrafast community \cite{MariyenkoOE2005,YamaneOE2012,BockOL2013,GrunwaldSR2014,YamaneNJoP2014}. Combining vortex pulses with high intensity ultrashort sources opens the way for strong-field physics with singular light beams \cite{ZurchNP2012,GariepyPRL2014}. Generating ultrashort vortices is challenging since most techniques used to create monochromatic vortices have limitations when applied to broadband laser sources \cite{MohAPL2006}. Even though polychromatic methods have been suggested and applied to ultrashort vortex generation \cite{MariyenkoOE2005,YamaneOE2012,AtenciaOE2013}, these are usually more complex to implement, making monochromatic techniques still attractive. 

An easy way to create ultrashort vortex fields is to use a spiral phase plate, e.g. made of glass. Transmission through a material with thickness \(L\) leads to an optical phase of \(\varphi(\lambda)=(2\pi c/\lambda) \Delta n(\lambda) L\), where $\lambda$ is the wavelength and $c$ the speed of light in vacuum, and $\Delta n$ the difference between the refractive index in the material and in vacuum. This phase varies with wavelength even in the absence of material dispersion (normal material dispersion makes the problem even worse). Therefore, the thickness jump of the spiral phase plate corresponds exactly to a phase equal to a multiple of \(2 \pi\) only at the design wavelength of the phase plate.
More generally the phase introduced by the glass plate varies according to:
\begin{equation}
    \varphi(\rho,\theta,\lambda)=\ell(\lambda)\theta.
\end{equation}
It is therefore not clear what impact the chromaticity of the phase plate has on the quality of the (broadband) ultrashort vortices.

Characterizing ultrashort vortex pulses is also challenging: a few-cycle optical vortex is an inherently three-dimensional object in space and time, so its complete characterization requires spatiotemporal characterization techniques. Knowing the spatiotemporal evolution of the electric field of optical vortices is important for simulating nonlinear effects driven by high-intensity vortex pulses, such as HHG \cite{Hernandez-GarciaPRL2013}, as the effect is directly dependent on the electric field.

In this work, we generate few-cycle optical vortex beams from an ultrafast titanium:sapphire (Ti:Sapph) laser oscillator using a spiral phase plate and characterize them in the spatiotemporal domain, using a technique based on spatially-resolved Fourier transform spectrometry \cite{MirandaOL2014,Gallet2014}. This technique is well suited for this particular case where high spatial resolution is needed, since scanning is performed only in one dimension (time). Our method allows for sub-cycle precision on the determination of relative phases across large areas.

\section{Experimental Setup}
\label{sec:experiment}

The experimental setup is schematically shown in Fig.~\ref{fig:setup}. Three main parts can be distinguished: generation of the vortex pulses, spatiotemporal measurement, where the vortex pulses are characterized by interfering them with a reference, and characterization of the reference pulse.
The ultrafast oscillator (Venteon Pulse-one) used as laser source for the experiments has a repetition rate of 80~MHz and around 3~nJ of energy per pulse with a spectrum supporting 5.0~fs pulses.
Chirped mirrors (Venteon DCM7) provide negative dispersion, and fused silica wedges are used for dispersion fine-tuning.
A 1~mm thick spiral phase plate (RPC Photonics) with topological charge of one at 800~nm, lithographically etched in a fused silica plate, is used in transmission to convert the laser output to a vortex beam.
The dispersion of the plate can easily be compensated for by removing glass insertion from the wedges. 
Unlike approaches based on polymeric phase plates or spatial light modulators, the etched phase plate can stand high intensity and high average power and is well suited for applications in nonlinear optics and strong-field physics.

%consequences this will have to the final vortex beam in the time domain, and how acceptable it is to use such technique with broadband lasers.}

The spatiotemporal characterization is performed using a technique based on spatially-resolved Fourier transform spectrometry \cite{MirandaOL2014}.
Part of the original beam is split before the spiral phase plate, sent through a delay stage, and used to generate a homogeneous spherical reference wave by focusing with an off-axis parabolic mirror to a pinhole for spatial filtering. 
The reference wave is collinearly recombined with the vortex beam and the resulting interference pattern is recorded as a function of delay with a CCD camera. Since the expanded reference beam is much larger than the original beam, we assume it to be homogeneous across the CCD array.
This way, an interferogram is obtained for each pixel, containing the spatially resolved phase information between the reference and the vortex beam. 
%If a reference beam was not available (i.e., if it was impossible to access the laser beam before turning it into a vortex), it could still be generated from the vortex pulse itself by filtering part of it, at the expense of energy losses.

\begin{figure}[htbp]
\centering
%\fbox{\includegraphics[]{d-scan.pdf}}
\includegraphics[]{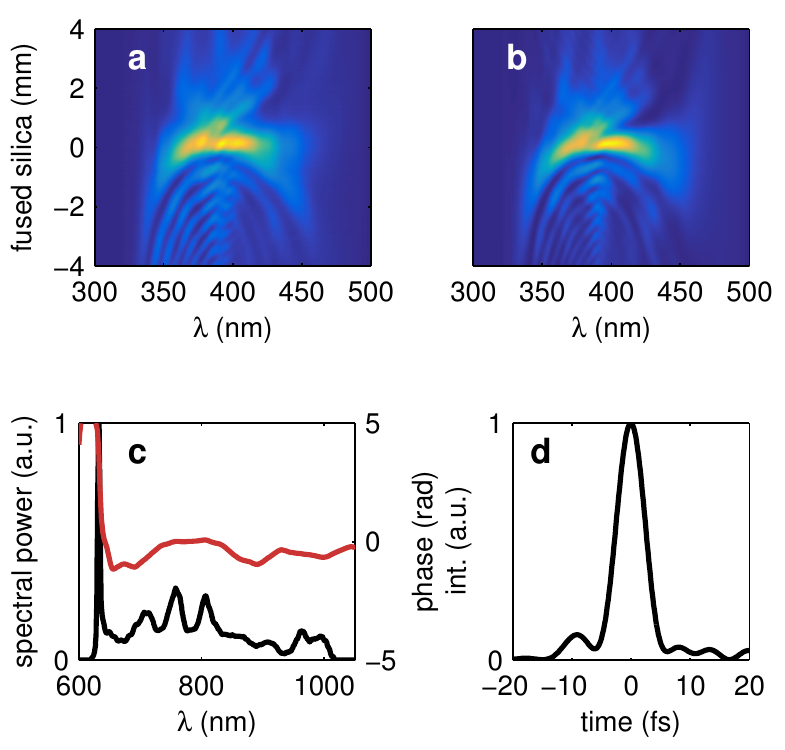}
\caption{Temporal characterization of the reference beam. (a) Measured and (b) retrieved  d-scan traces, (c) measured fundamental spectrum (black line) and retrieved phase (red line) and (d) reconstructed temporal profile at the reference glass insertion (zero amount of glass on the d-scan plots). The pulse duration (FWHM) at the reference insertion is 5.5~fs FWHM.}
\label{fig:d-scan}
\end{figure}

The temporal characterization of the reference wave is performed using the d-scan technique \cite{MirandaOE2012}.
A small central portion of the beam is selected by an iris with about \(200~\mu m\) diameter (see Fig.~\ref{fig:setup}). This portion of the beam contains sufficient power for second harmonic generation (SHG), while being small enough to be assumed spatially homogeneous. An off-axis parabolic mirror focuses the beam to a thin (\(20~\mu m\))  KDP crystal and a broadband polarizer blocks the fundamental NIR light from entering the fiber that sends the SHG signal to a spectrometer. The SHG spectrum is recorded as function of glass insertion (dispersion) around the optimal compression insertion. From the SHG trace and the measured fundamental spectrum, the spectral phase (Fig.~\ref{fig:d-scan}(c)) is retrieved using an iterative numeric algorithm \cite{MirandaOE2012}. From the retrieved phase, the corresponding time domain pulse can be obtained at any glass insertion.
The shortest pulses obtained have a FWHM duration of 5.5~fs (Fig.~\ref{fig:d-scan}(d)).
We assume that the measured spectral phase is a good approximation to that of the spherical reference pulse used in the main interferometer setup, which is used for the spatiotemporal reconstruction presented below.

\section{Spatially Resolved Fourier Transform Spectrometry} 
\label{sec:theory}
We first explain the principle of Fourier transform spectroscopy considering only one point in space, thus omitting the spatial coordinates. Consider an ultrashort pulse, described by the complex field \(U(t)=|U(t)|\exp [i\psi(t)]\), or its corresponding complex spectrum \(\tilde{U}(\omega)=|\tilde{U}(\omega)|\exp[i\phi(\omega)]\), interfering with a reference pulse \(U_r(t)=|U_r(t)|\exp [i\psi_r(t)]\). 
The obtained linear interferogram, measured with an intensity detector as a function of delay, is given by
\begin{equation}
I(\tau)=\int\left|U(t)+U_r(t-\tau)\right|^2 dt.
\end{equation}
The Fourier transform of this interferogram can be written as:
\begin{eqnarray}
	\label{eq:fourier_interferogram}
	\F{I(\tau)}	&=&\F{\int\left|U(t)\right|^2 dt + \int\left|U_r(t)\right|^2 dt}  \nonumber \\ 
	       	&+& \tilde{U}(\omega)\tilde{U}_r^*(\omega) \nonumber \\ &+&\tilde{U}^*(-\omega)\tilde{U}_r(-\omega).
\end{eqnarray}
Equation~\ref{eq:fourier_interferogram} is composed of three terms, the first one being the DC component of the interferogram and the second and third terms being the complex conjugate of each other.
The three terms are easily separable in the frequency domain.
We choose to isolate the second term:
\begin{eqnarray}
	\tilde{A}(\omega)&=&\tilde{U}(\omega)\tilde{U}_r^*(\omega) \nonumber \\ &=&|\tilde{U}(\omega)||\tilde{U}_r(\omega)|\exp[i\phi(\omega)-i\phi_r(\omega)].
\end{eqnarray}

If the reference phase \(\phi_r(\omega)\) is known and the reference spectrum \(|\tilde{U}_r(\omega)|\) covers the bandwidth of the field to be measured, then \(\tilde{U}(\omega)\) can be retrieved:
\begin{equation}
\tilde{U}(\omega)=\frac{\tilde{A}(\omega)}{|\tilde{U}_r(\omega)|}\exp[i\phi_r(\omega)].
\end{equation}

We now consider the spatial transverse dependence of the waves. If the reference wave is homogeneous, \(\tilde{U}_r(x,y,\omega)=\tilde{U}_r(\omega)\), the spatially resolved spectral amplitude and field of the unknown pulse can be found:
\begin{equation}
	\tilde{U}(x,y,\omega)=\frac{\tilde{A}(x,y,\omega)}{|\tilde{U}_r(\omega)|}\exp[i\phi_r(\omega)],
\end{equation}
as well as the corresponding time domain field \(U(x,y,t)\), by Fourier transformation.
In practice, it is much easier to produce a spherical reference wave (Fig.~\ref{fig:setup}). The spherical curvature of the reference wave can easily be removed numerically from the measured interferogram.

It should be noted that the spatial phase front can be extracted for every wavelength even without knowledge of the spectral phase of the reference. However, the complete spectral phase and thus the pulse duration of the unknown pulse cannot be obtained without knowledge of the reference pulse.

\section{Frequency-Dependent Spatial Characterization}
\label{sec:freq_charac}

\begin{figure}[htbp]
\centering
%\fbox{\includegraphics[]{profiles.pdf}}
\includegraphics[]{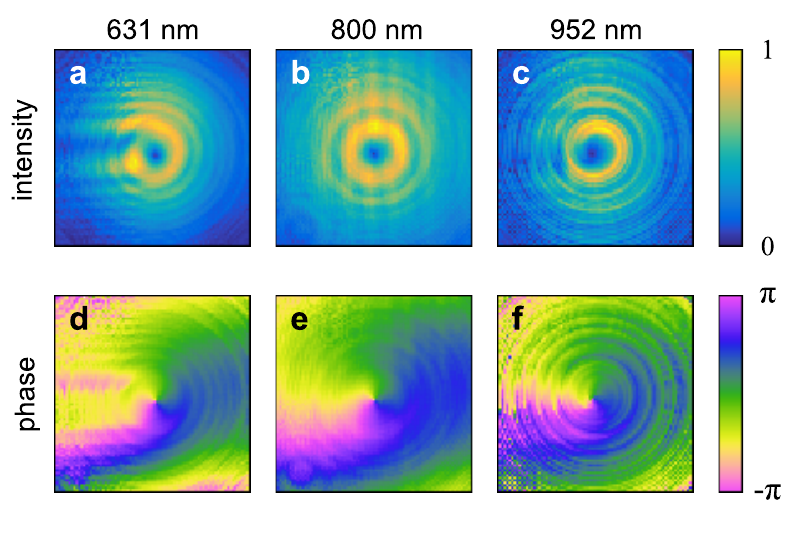}
\caption{(a-c) Intensity and (d-f) phase profiles  for three different wavelengths. The dimensions of each of the plots are 4~mm x 4~mm.}
\label{fig:profiles}
\end{figure}

From the complete spatiotemporally characterized field \(\tilde{U}(x,y,\omega)\) the frequency-dependent characteristics of the vortex can be investigated.
For any given wavelength \(\lambda_a \) (or corresponding frequency \(\omega_a\)) the complex 2D array \(\tilde{U}(x,y,\omega_a)\) can be selected and the intensity (the absolute square of the amplitude) and phase can be obtained.

Fig.~\ref{fig:profiles} shows intensity and phase profiles at the CCD detector for three different wavelengths. 
For the design wavelength of the phase plate (800~nm) a clean vortex is obtained.
The effect of the plate's chromaticity increases as the wavelength departs from the nominal one.

At 631~nm (Fig.~\ref{fig:profiles}(a)) strong diffraction effects, giving rise to horizontal stripes, are noticeable. They originate from the phase jump introduced by the plate (\(\approx 2.5 \pi\) at 631~nm).
At longer wavelengths the effect is not as clearly pronounced, mostly because material dispersion is lower making the phase jump closer to \(2 \pi\). Figures~\ref{fig:profiles}(c,f) show the reconstructed profile at 952~nm. At this wavelength, the phase jump is estimated to be around \(1.7 \pi\), leading to less evident distortion on the intensity and phase profiles. We find that for any wavelength within the spectrum of the pulse, there is a phase singularity present in the center of the beam with the intensity dropping to zero.

\begin{figure}[htbp]
\centering
%\fbox{\includegraphics[]{profiles.pdf}}
\includegraphics[]{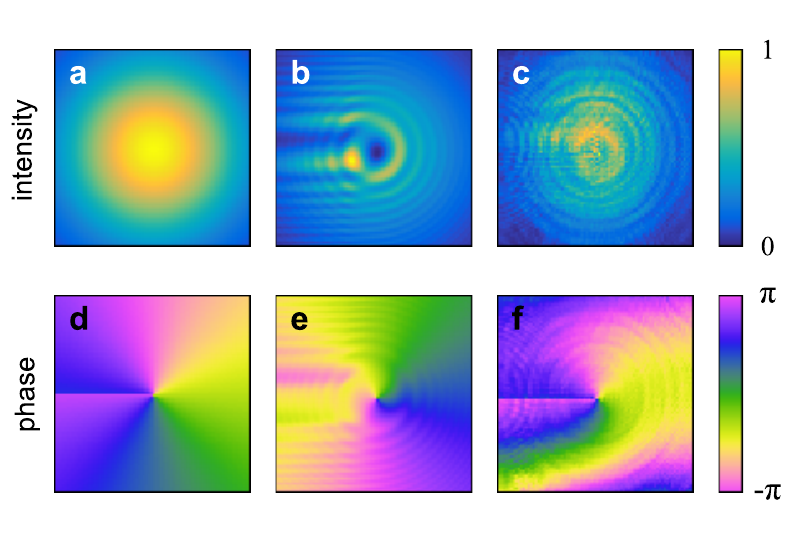}
\caption{Simulated intensity and phase profiles (a,d) after the spiral phase plate  and (b,e) numerically propagated to the position of the detector. We assume a Gaussian beam and a phase step of \(2.5\pi\) due to the plate. (c,f) Calculated intensity and phase profiles obtained by back-propagating the measured field (Fig.~\ref{fig:profiles}(a,d)) to the plane of the plate. The dimensions of each of the plots are 4~mm x 4~mm.}
\label{fig:sim_profiles}
\end{figure}

The effects originating from phase plates with  phase jumps not equal to \(2 \pi\) (see e.g. \cite{BerryJoOAPaAO2004}) can be reproduced by simple propagation simulations.
Figure~\ref{fig:sim_profiles} shows an optical field, originally Gaussian, just after a spiral phase plate with a phase jump of \(2.5\pi\) (Fig.~\ref{fig:sim_profiles}(a,d)) and after propagation to the position of the detector (Fig.~\ref{fig:sim_profiles}(b,e)).
The simulated intensity and phase profiles (Fig.~\ref{fig:sim_profiles}(b,e)) qualitatively agree with the measured profiles at 631~nm (Fig.~\ref{fig:profiles}(a,d)).
The differences between measured and simulated phase profiles away from the center are mostly due to astigmatism which was not accounted for in the simulation.

We also back-propagated the measured field (Fig.~\ref{fig:profiles}(a,d)) to the plane containing the phase plate. The results are plotted in Fig.~\ref{fig:sim_profiles}(c,f), clearly showing a phase discontinuity (around \(0.5\pi\)) due to the thickness jump of the phase-plate, as well as a slight astigmatism of the used beam.

\section{Time-Dependent Spatial Characterization}

\begin{figure*}[htbp]
\centering
%\fbox{\includegraphics[]{3d test.pdf}}
\includegraphics[]{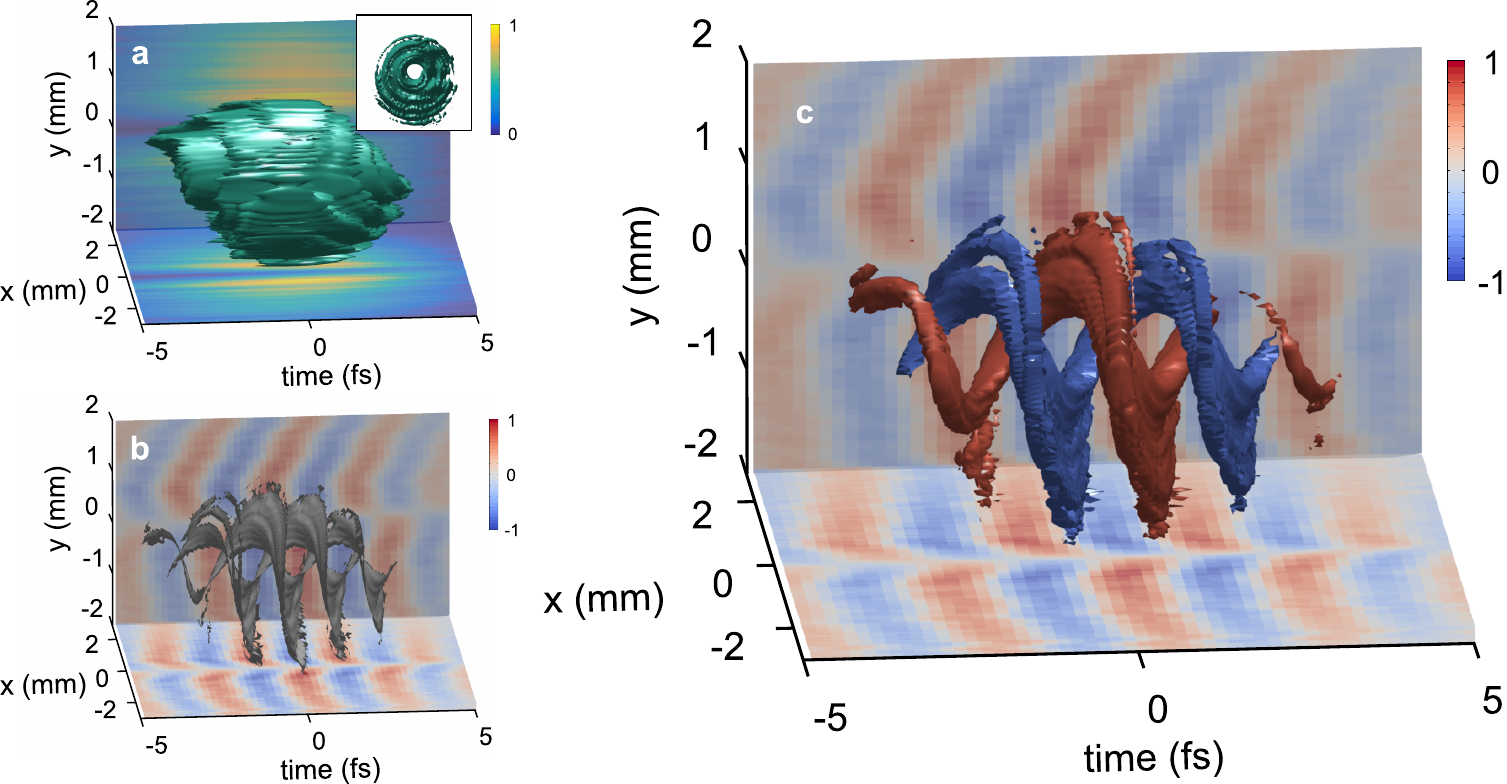}
\caption{Three dimensional reconstruction of the vortex beam.  (a) Intensity profile in space and time obtained by taking the square of the absolute field. The plot shows an isosurface set at half the peak intensity. The inset shows the same surface from a view along the propagation direction, visualizing the phase singularity (dimensions of inset plot are 4~mm x 4~mm). (b) Isosurface set at 0 of the real part of the complex field. For clarity, only regions where the intensity is higher than 0.4 times the peak intensity are shown. (c) Isosurfaces of the real part of the field, set at 0.7 (red) and -0.7 (blue) of the peak field. }
\label{fig:3d_plot}
\end{figure*}

From the complete description in the spectral domain \(\tilde{U}(x,y,\omega)\) the spatiotemporal field \(U(x,y,t)\) can be obtained by simply Fourier transforming at each point in space. To achieve the shortest possible vortices as shown here, 1~mm of fused silica (corresponding to the substrate of the spiral phase plate) was removed by adjusting the wedges used for compression. 

Figure~\ref{fig:3d_plot}(a) shows the spatiotemporal intensity profile \(|U(x,y,t)|^2\) as an isosurface set at half the peak intensity value. The projections on the vertical and horizontal planes in Fig.~\ref{fig:3d_plot} are obtained by taking cuts through the center of the beam.
The time duration of the pulse taken at any individual point on the main ring is approximately the same as for the reference pulse, i.e. \(\approx\) 5.5~fs.
However, considering the vortex as a whole, the duration is longer, around 8~fs.
The largest contribution to that is the phase plate itself, which delays parts of the beam with respect to others. Furthermore, residual astigmatism and diffraction effects resulting from the chromaticity of the phase plate contribute to the longer pulse duration.
The phase singularity, apparent as vanishing intensity in the center of the beam, can clearly be observed (inset of Fig.~\ref{fig:3d_plot}(a)).

To visualize the spiraling nature of the electric field in time we take the real part of the complex field, \(E(x,y,t)=\Re\{U(x,y,t)\}\).
Figure~\ref{fig:3d_plot}(b) shows isosurfaces of \(E(x,y,t)\) set at a value of zero, which shows the wavefront of the spiraling field, while  Fig.~\ref{fig:3d_plot}(c) shows isosurfaces set at 0.7 and -0.7 of the peak field value, respectively.
Note that the absolute phase of the electric field is arbitrary, since the carrier-to-envelope offset phase of the reference was not measured.

\section{Conclusions}

We have demonstrated the generation and complete characterization of ultrashort vortex pulses, using simple to implement yet powerful techniques.

The combination of the d-scan technique with spatially resolved Fourier transform interferometry has proven to be a powerful technique for the characterization of ultrashort optical pulses with complex spatial distribution and spatiotemporal couplings, being especially suited for measurements requiring high spatial resolution with accurate phase sensitivity across the whole beam profile.

The spiral phase plate, although not designed for broadband operation, has a good trade off between performance and ease of use, since it is extremely simple to align, has low losses, and is well suited for high-power applications.
Despite the chromatic response of the spiral phase plate, which leads to a break of circular symmetry for spectral portions of the pulse away from the design wavelength, the phase singularity in the center of the beam remains intact. 

Our characterization shows that high power ultrashort vortex pulses can conveniently be generated with spiral glass phase plates. This method should be useful for strong field physics with singular light fields.

\section*{Funding Information}
This work was partly supported by the European Research Council (PALP and CLIAS), the Knut and Alice Wallenberg Foundation, the Joint Research Programme INREX of Laserlab-Europe III, and the Swedish Research Council.


\begin{thebibliography}{26}%
\makeatletter
\providecommand \@ifxundefined [1]{%
 \@ifx{#1\undefined}
}%
\providecommand \@ifnum [1]{%
 \ifnum #1\expandafter \@firstoftwo
 \else \expandafter \@secondoftwo
 \fi
}%
\providecommand \@ifx [1]{%
 \ifx #1\expandafter \@firstoftwo
 \else \expandafter \@secondoftwo
 \fi
}%
\providecommand \natexlab [1]{#1}%
\providecommand \enquote  [1]{``#1''}%
\providecommand \bibnamefont  [1]{#1}%
\providecommand \bibfnamefont [1]{#1}%
\providecommand \citenamefont [1]{#1}%
\providecommand \href@noop [0]{\@secondoftwo}%
\providecommand \href [0]{\begingroup \@sanitize@url \@href}%
\providecommand \@href[1]{\@@startlink{#1}\@@href}%
\providecommand \@@href[1]{\endgroup#1\@@endlink}%
\providecommand \@sanitize@url [0]{\catcode `\\12\catcode `\$12\catcode
  `\&12\catcode `\#12\catcode `\^12\catcode `\_12\catcode `\%12\relax}%
\providecommand \@@startlink[1]{}%
\providecommand \@@endlink[0]{}%
\providecommand \url  [0]{\begingroup\@sanitize@url \@url }%
\providecommand \@url [1]{\endgroup\@href {#1}{\urlprefix }}%
\providecommand \urlprefix  [0]{URL }%
\providecommand \Eprint [0]{\href }%
\providecommand \doibase [0]{http://dx.doi.org/}%
\providecommand \selectlanguage [0]{\@gobble}%
\providecommand \bibinfo  [0]{\@secondoftwo}%
\providecommand \bibfield  [0]{\@secondoftwo}%
\providecommand \translation [1]{[#1]}%
\providecommand \BibitemOpen [0]{}%
\providecommand \bibitemStop [0]{}%
\providecommand \bibitemNoStop [0]{.\EOS\space}%
\providecommand \EOS [0]{\spacefactor3000\relax}%
\providecommand \BibitemShut  [1]{\csname bibitem#1\endcsname}%
\let\auto@bib@innerbib\@empty
%</preamble>
\bibitem [{\citenamefont {Nye}\ and\ \citenamefont
  {Berry}(1974)}]{NyePotRSoLAMPaES1974}%
  \BibitemOpen
  \bibfield  {author} {\bibinfo {author} {\bibfnamefont {J.~F.}\ \bibnamefont
  {Nye}}\ and\ \bibinfo {author} {\bibfnamefont {M.~V.}\ \bibnamefont
  {Berry}},\ }\href {\doibase 10.1098/rspa.1974.0012} {\bibfield  {journal}
  {\bibinfo  {journal} {Proceedings of the Royal Society of London A:
  Mathematical, Physical and Engineering Sciences P. Roy. Soc. Lond. A: Mat.}\
  }\textbf {\bibinfo {volume} {336}},\ \bibinfo {pages} {165} (\bibinfo {year}
  {1974})}\BibitemShut {NoStop}%
\bibitem [{\citenamefont {Coullet}\ \emph {et~al.}(1989)\citenamefont
  {Coullet}, \citenamefont {Gil},\ and\ \citenamefont {Rocca}}]{CoulletOC1989}%
  \BibitemOpen
  \bibfield  {author} {\bibinfo {author} {\bibfnamefont {P.}~\bibnamefont
  {Coullet}}, \bibinfo {author} {\bibfnamefont {L.}~\bibnamefont {Gil}}, \ and\
  \bibinfo {author} {\bibfnamefont {F.}~\bibnamefont {Rocca}},\ }\href
  {\doibase http://dx.doi.org/10.1016/0030-4018(89)90180-6} {\bibfield
  {journal} {\bibinfo  {journal} {Optics Communications}\ }\textbf {\bibinfo
  {volume} {73}},\ \bibinfo {pages} {403 } (\bibinfo {year}
  {1989})}\BibitemShut {NoStop}%
\bibitem [{\citenamefont {Allen}\ \emph {et~al.}(1992)\citenamefont {Allen},
  \citenamefont {Beijersbergen}, \citenamefont {Spreeuw},\ and\ \citenamefont
  {Woerdman}}]{AllenPRA1992}%
  \BibitemOpen
  \bibfield  {author} {\bibinfo {author} {\bibfnamefont {L.}~\bibnamefont
  {Allen}}, \bibinfo {author} {\bibfnamefont {M.}~\bibnamefont
  {Beijersbergen}}, \bibinfo {author} {\bibfnamefont {R.}~\bibnamefont
  {Spreeuw}}, \ and\ \bibinfo {author} {\bibfnamefont {J.}~\bibnamefont
  {Woerdman}},\ }\href {\doibase 10.1103/PhysRevA.45.8185} {\bibfield
  {journal} {\bibinfo  {journal} {Phys. Rev. A}\ }\textbf {\bibinfo {volume}
  {45}},\ \bibinfo {pages} {8185} (\bibinfo {year} {1992})}\BibitemShut
  {NoStop}%
\bibitem [{\citenamefont {Allen}\ \emph {et~al.}(1999)\citenamefont {Allen},
  \citenamefont {Padgett},\ and\ \citenamefont {Babiker}}]{Allen1999}%
  \BibitemOpen
  \bibfield  {author} {\bibinfo {author} {\bibfnamefont {L.}~\bibnamefont
  {Allen}}, \bibinfo {author} {\bibfnamefont {M.}~\bibnamefont {Padgett}}, \
  and\ \bibinfo {author} {\bibfnamefont {M.}~\bibnamefont {Babiker}}\
  }(\bibinfo  {publisher} {Elsevier},\ \bibinfo {year} {1999})\ pp.\ \bibinfo
  {pages} {291 -- 372}\BibitemShut {NoStop}%
\bibitem [{\citenamefont {Yao}\ and\ \citenamefont
  {Padgett}(2011)}]{YaoAOP2011}%
  \BibitemOpen
  \bibfield  {author} {\bibinfo {author} {\bibfnamefont {A.~M.}\ \bibnamefont
  {Yao}}\ and\ \bibinfo {author} {\bibfnamefont {M.~J.}\ \bibnamefont
  {Padgett}},\ }\href {\doibase 10.1364/AOP.3.000161} {\bibfield  {journal}
  {\bibinfo  {journal} {Adv. Opt. Photon.}\ }\textbf {\bibinfo {volume} {3}},\
  \bibinfo {pages} {161} (\bibinfo {year} {2011})}\BibitemShut {NoStop}%
\bibitem [{\citenamefont {Swartzlander}(2001)}]{SwartzlanderOL2001}%
  \BibitemOpen
  \bibfield  {author} {\bibinfo {author} {\bibfnamefont {G.~A.}\ \bibnamefont
  {Swartzlander}, \bibfnamefont {Jr.}},\ }\href {\doibase 10.1364/OL.26.000497}
  {\bibfield  {journal} {\bibinfo  {journal} {Opt. Lett.}\ }\textbf {\bibinfo
  {volume} {26}},\ \bibinfo {pages} {497} (\bibinfo {year} {2001})}\BibitemShut
  {NoStop}%
\bibitem [{\citenamefont {Bretschneider}\ \emph {et~al.}(2007)\citenamefont
  {Bretschneider}, \citenamefont {Eggeling},\ and\ \citenamefont
  {Hell}}]{BretschneiderPRL2007}%
  \BibitemOpen
  \bibfield  {author} {\bibinfo {author} {\bibfnamefont {S.}~\bibnamefont
  {Bretschneider}}, \bibinfo {author} {\bibfnamefont {C.}~\bibnamefont
  {Eggeling}}, \ and\ \bibinfo {author} {\bibfnamefont {S.~W.}\ \bibnamefont
  {Hell}},\ }\href {\doibase 10.1103/PhysRevLett.98.218103} {\bibfield
  {journal} {\bibinfo  {journal} {Phys. Rev. Lett.}\ }\textbf {\bibinfo
  {volume} {98}},\ \bibinfo {pages} {218103} (\bibinfo {year}
  {2007})}\BibitemShut {NoStop}%
\bibitem [{\citenamefont {O'Neil}\ and\ \citenamefont
  {Padgett}(2000)}]{ONeilOC2000}%
  \BibitemOpen
  \bibfield  {author} {\bibinfo {author} {\bibfnamefont {A.~T.}\ \bibnamefont
  {O'Neil}}\ and\ \bibinfo {author} {\bibfnamefont {M.~J.}\ \bibnamefont
  {Padgett}},\ }\href {\doibase
  http://dx.doi.org/10.1016/S0030-4018(00)00989-5} {\bibfield  {journal}
  {\bibinfo  {journal} {Opt. Commun.}\ }\textbf {\bibinfo {volume} {185}},\
  \bibinfo {pages} {139 } (\bibinfo {year} {2000})}\BibitemShut {NoStop}%
\bibitem [{\citenamefont {Beijersbergen}\ \emph {et~al.}(1994)\citenamefont
  {Beijersbergen}, \citenamefont {Coerwinkel}, \citenamefont {Kristensen},\
  and\ \citenamefont {Woerdman}}]{BeijersbergenOC1994}%
  \BibitemOpen
  \bibfield  {author} {\bibinfo {author} {\bibfnamefont {M.}~\bibnamefont
  {Beijersbergen}}, \bibinfo {author} {\bibfnamefont {R.}~\bibnamefont
  {Coerwinkel}}, \bibinfo {author} {\bibfnamefont {M.}~\bibnamefont
  {Kristensen}}, \ and\ \bibinfo {author} {\bibfnamefont {J.}~\bibnamefont
  {Woerdman}},\ }\href {\doibase
  http://dx.doi.org/10.1016/0030-4018(94)90638-6} {\bibfield  {journal}
  {\bibinfo  {journal} {Opt. Commun.}\ }\textbf {\bibinfo {volume} {112}},\
  \bibinfo {pages} {321 } (\bibinfo {year} {1994})}\BibitemShut {NoStop}%
\bibitem [{\citenamefont {Bazhenov}\ \emph {et~al.}(1992)\citenamefont
  {Bazhenov}, \citenamefont {Soskin},\ and\ \citenamefont
  {Vasnetsov}}]{BazhenovJoMO1992}%
  \BibitemOpen
  \bibfield  {author} {\bibinfo {author} {\bibfnamefont {V.}~\bibnamefont
  {Bazhenov}}, \bibinfo {author} {\bibfnamefont {M.}~\bibnamefont {Soskin}}, \
  and\ \bibinfo {author} {\bibfnamefont {M.}~\bibnamefont {Vasnetsov}},\ }\href
  {\doibase 10.1080/09500349214551011} {\bibfield  {journal} {\bibinfo
  {journal} {J. Mod. Opt.}\ }\textbf {\bibinfo {volume} {39}},\ \bibinfo
  {pages} {985} (\bibinfo {year} {1992})},\ \Eprint
  {http://arxiv.org/abs/http://dx.doi.org/10.1080/09500349214551011}
  {http://dx.doi.org/10.1080/09500349214551011} \BibitemShut {NoStop}%
\bibitem [{\citenamefont {Bock}\ \emph {et~al.}(2012)\citenamefont {Bock},
  \citenamefont {Jahns},\ and\ \citenamefont {Grunwald}}]{BockOL2012}%
  \BibitemOpen
  \bibfield  {author} {\bibinfo {author} {\bibfnamefont {M.}~\bibnamefont
  {Bock}}, \bibinfo {author} {\bibfnamefont {J.}~\bibnamefont {Jahns}}, \ and\
  \bibinfo {author} {\bibfnamefont {R.}~\bibnamefont {Grunwald}},\ }\href
  {\doibase 10.1364/OL.37.003804} {\bibfield  {journal} {\bibinfo  {journal}
  {Opt. Lett.}\ }\textbf {\bibinfo {volume} {37}},\ \bibinfo {pages} {3804}
  (\bibinfo {year} {2012})}\BibitemShut {NoStop}%
\bibitem [{\citenamefont {Musigmann}\ \emph {et~al.}(2014)\citenamefont
  {Musigmann}, \citenamefont {Jahns}, \citenamefont {Bock},\ and\ \citenamefont
  {Grunwald}}]{MusigmannAO2014}%
  \BibitemOpen
  \bibfield  {author} {\bibinfo {author} {\bibfnamefont {M.}~\bibnamefont
  {Musigmann}}, \bibinfo {author} {\bibfnamefont {J.}~\bibnamefont {Jahns}},
  \bibinfo {author} {\bibfnamefont {M.}~\bibnamefont {Bock}}, \ and\ \bibinfo
  {author} {\bibfnamefont {R.}~\bibnamefont {Grunwald}},\ }\href {\doibase
  10.1364/AO.53.007304} {\bibfield  {journal} {\bibinfo  {journal} {Appl.
  Opt.}\ }\textbf {\bibinfo {volume} {53}},\ \bibinfo {pages} {7304} (\bibinfo
  {year} {2014})}\BibitemShut {NoStop}%
\bibitem [{\citenamefont {Mariyenko}\ \emph {et~al.}(2005)\citenamefont
  {Mariyenko}, \citenamefont {Strohaber},\ and\ \citenamefont
  {Uiterwaal}}]{MariyenkoOE2005}%
  \BibitemOpen
  \bibfield  {author} {\bibinfo {author} {\bibfnamefont {I.}~\bibnamefont
  {Mariyenko}}, \bibinfo {author} {\bibfnamefont {J.}~\bibnamefont
  {Strohaber}}, \ and\ \bibinfo {author} {\bibfnamefont {C.}~\bibnamefont
  {Uiterwaal}},\ }\href {\doibase 10.1364/OPEX.13.007599} {\bibfield  {journal}
  {\bibinfo  {journal} {Opt. Express}\ }\textbf {\bibinfo {volume} {13}},\
  \bibinfo {pages} {7599} (\bibinfo {year} {2005})}\BibitemShut {NoStop}%
\bibitem [{\citenamefont {Yamane}\ \emph {et~al.}(2012)\citenamefont {Yamane},
  \citenamefont {Toda},\ and\ \citenamefont {Morita}}]{YamaneOE2012}%
  \BibitemOpen
  \bibfield  {author} {\bibinfo {author} {\bibfnamefont {K.}~\bibnamefont
  {Yamane}}, \bibinfo {author} {\bibfnamefont {Y.}~\bibnamefont {Toda}}, \ and\
  \bibinfo {author} {\bibfnamefont {R.}~\bibnamefont {Morita}},\ }\href
  {\doibase 10.1364/OE.20.018986} {\bibfield  {journal} {\bibinfo  {journal}
  {Opt. Express}\ }\textbf {\bibinfo {volume} {20}},\ \bibinfo {pages} {18986}
  (\bibinfo {year} {2012})}\BibitemShut {NoStop}%
\bibitem [{\citenamefont {Bock}\ \emph {et~al.}(2013)\citenamefont {Bock},
  \citenamefont {Brunne}, \citenamefont {Treffer}, \citenamefont {K\"{o}nig},
  \citenamefont {Wallrabe},\ and\ \citenamefont {Grunwald}}]{BockOL2013}%
  \BibitemOpen
  \bibfield  {author} {\bibinfo {author} {\bibfnamefont {M.}~\bibnamefont
  {Bock}}, \bibinfo {author} {\bibfnamefont {J.}~\bibnamefont {Brunne}},
  \bibinfo {author} {\bibfnamefont {A.}~\bibnamefont {Treffer}}, \bibinfo
  {author} {\bibfnamefont {S.}~\bibnamefont {K\"{o}nig}}, \bibinfo {author}
  {\bibfnamefont {U.}~\bibnamefont {Wallrabe}}, \ and\ \bibinfo {author}
  {\bibfnamefont {R.}~\bibnamefont {Grunwald}},\ }\href {\doibase
  10.1364/OL.38.003642} {\bibfield  {journal} {\bibinfo  {journal} {Opt.
  Lett.}\ }\textbf {\bibinfo {volume} {38}},\ \bibinfo {pages} {3642} (\bibinfo
  {year} {2013})}\BibitemShut {NoStop}%
\bibitem [{\citenamefont {Grunwald}\ \emph {et~al.}(2014)\citenamefont
  {Grunwald}, \citenamefont {Elsaesser},\ and\ \citenamefont
  {Bock}}]{GrunwaldSR2014}%
  \BibitemOpen
  \bibfield  {author} {\bibinfo {author} {\bibfnamefont {R.}~\bibnamefont
  {Grunwald}}, \bibinfo {author} {\bibfnamefont {T.}~\bibnamefont {Elsaesser}},
  \ and\ \bibinfo {author} {\bibfnamefont {M.}~\bibnamefont {Bock}},\ }\href
  {http://dx.doi.org/10.1038/srep07148} {\bibfield  {journal} {\bibinfo
  {journal} {Sci. Rep.}\ }\textbf {\bibinfo {volume} {4}} (\bibinfo {year}
  {2014})}\BibitemShut {NoStop}%
\bibitem [{\citenamefont {Yamane}\ \emph {et~al.}(2014)\citenamefont {Yamane},
  \citenamefont {Yang}, \citenamefont {Toda},\ and\ \citenamefont
  {Morita}}]{YamaneNJoP2014}%
  \BibitemOpen
  \bibfield  {author} {\bibinfo {author} {\bibfnamefont {K.}~\bibnamefont
  {Yamane}}, \bibinfo {author} {\bibfnamefont {Z.}~\bibnamefont {Yang}},
  \bibinfo {author} {\bibfnamefont {Y.}~\bibnamefont {Toda}}, \ and\ \bibinfo
  {author} {\bibfnamefont {R.}~\bibnamefont {Morita}},\ }\href
  {http://stacks.iop.org/1367-2630/16/i=5/a=053020} {\bibfield  {journal}
  {\bibinfo  {journal} {New J. Phys.}\ }\textbf {\bibinfo {volume} {16}},\
  \bibinfo {pages} {053020} (\bibinfo {year} {2014})}\BibitemShut {NoStop}%
\bibitem [{\citenamefont {Z\"{u}rch}\ \emph {et~al.}(2012)\citenamefont
  {Z\"{u}rch}, \citenamefont {Kern}, \citenamefont {Hansinger}, \citenamefont
  {Dreischuh},\ and\ \citenamefont {Spielmann}}]{ZurchNP2012}%
  \BibitemOpen
  \bibfield  {author} {\bibinfo {author} {\bibfnamefont {M.}~\bibnamefont
  {Z\"{u}rch}}, \bibinfo {author} {\bibfnamefont {C.}~\bibnamefont {Kern}},
  \bibinfo {author} {\bibfnamefont {P.}~\bibnamefont {Hansinger}}, \bibinfo
  {author} {\bibfnamefont {A.}~\bibnamefont {Dreischuh}}, \ and\ \bibinfo
  {author} {\bibfnamefont {C.}~\bibnamefont {Spielmann}},\ }\href
  {http://dx.doi.org/10.1038/nphys2397} {\bibfield  {journal} {\bibinfo
  {journal} {Nat. Phys.}\ }\textbf {\bibinfo {volume} {8}},\ \bibinfo {pages}
  {743} (\bibinfo {year} {2012})}\BibitemShut {NoStop}%
\bibitem [{\citenamefont {Gariepy}\ \emph {et~al.}(2014)\citenamefont
  {Gariepy}, \citenamefont {Leach}, \citenamefont {Kim}, \citenamefont
  {Hammond}, \citenamefont {Frumker}, \citenamefont {Boyd},\ and\ \citenamefont
  {Corkum}}]{GariepyPRL2014}%
  \BibitemOpen
  \bibfield  {author} {\bibinfo {author} {\bibfnamefont {G.}~\bibnamefont
  {Gariepy}}, \bibinfo {author} {\bibfnamefont {J.}~\bibnamefont {Leach}},
  \bibinfo {author} {\bibfnamefont {K.~T.}\ \bibnamefont {Kim}}, \bibinfo
  {author} {\bibfnamefont {T.~J.}\ \bibnamefont {Hammond}}, \bibinfo {author}
  {\bibfnamefont {E.}~\bibnamefont {Frumker}}, \bibinfo {author} {\bibfnamefont
  {R.~W.}\ \bibnamefont {Boyd}}, \ and\ \bibinfo {author} {\bibfnamefont
  {P.~B.}\ \bibnamefont {Corkum}},\ }\href {\doibase
  10.1103/PhysRevLett.113.153901} {\bibfield  {journal} {\bibinfo  {journal}
  {Phys. Rev. Lett.}\ }\textbf {\bibinfo {volume} {113}},\ \bibinfo {pages}
  {153901} (\bibinfo {year} {2014})}\BibitemShut {NoStop}%
\bibitem [{\citenamefont {Moh}\ \emph {et~al.}(2006)\citenamefont {Moh},
  \citenamefont {Yuan}, \citenamefont {Tang}, \citenamefont {Cheong},
  \citenamefont {Zhang}, \citenamefont {Low}, \citenamefont {Peng},
  \citenamefont {Niu},\ and\ \citenamefont {Lin}}]{MohAPL2006}%
  \BibitemOpen
  \bibfield  {author} {\bibinfo {author} {\bibfnamefont {K.~J.}\ \bibnamefont
  {Moh}}, \bibinfo {author} {\bibfnamefont {X.-C.}\ \bibnamefont {Yuan}},
  \bibinfo {author} {\bibfnamefont {D.~Y.}\ \bibnamefont {Tang}}, \bibinfo
  {author} {\bibfnamefont {W.~C.}\ \bibnamefont {Cheong}}, \bibinfo {author}
  {\bibfnamefont {L.~S.}\ \bibnamefont {Zhang}}, \bibinfo {author}
  {\bibfnamefont {D.~K.~Y.}\ \bibnamefont {Low}}, \bibinfo {author}
  {\bibfnamefont {X.}~\bibnamefont {Peng}}, \bibinfo {author} {\bibfnamefont
  {H.~B.}\ \bibnamefont {Niu}}, \ and\ \bibinfo {author} {\bibfnamefont
  {Z.~Y.}\ \bibnamefont {Lin}},\ }\href {\doibase
  http://dx.doi.org/10.1063/1.2178507} {\bibfield  {journal} {\bibinfo
  {journal} {Appl. Phys. Lett.}\ }\textbf {\bibinfo {volume} {88}},\ \bibinfo
  {eid} {091103} (\bibinfo {year} {2006}),\
  http://dx.doi.org/10.1063/1.2178507}\BibitemShut {NoStop}%
\bibitem [{\citenamefont {Atencia}\ \emph {et~al.}(2013)\citenamefont
  {Atencia}, \citenamefont {Collados}, \citenamefont {Quintanilla},
  \citenamefont {Mar{\'i}n-S{\'a}ez},\ and\ \citenamefont
  {Sola}}]{AtenciaOE2013}%
  \BibitemOpen
  \bibfield  {author} {\bibinfo {author} {\bibfnamefont {J.}~\bibnamefont
  {Atencia}}, \bibinfo {author} {\bibfnamefont {M.-V.}\ \bibnamefont
  {Collados}}, \bibinfo {author} {\bibfnamefont {M.}~\bibnamefont
  {Quintanilla}}, \bibinfo {author} {\bibfnamefont {J.}~\bibnamefont
  {Mar{\'i}n-S{\'a}ez}}, \ and\ \bibinfo {author} {\bibfnamefont {{\'I}.~J.}\
  \bibnamefont {Sola}},\ }\href {\doibase 10.1364/OE.21.021056} {\bibfield
  {journal} {\bibinfo  {journal} {Opt. Express}\ }\textbf {\bibinfo {volume}
  {21}},\ \bibinfo {pages} {21056} (\bibinfo {year} {2013})}\BibitemShut
  {NoStop}%
\bibitem [{\citenamefont {Hern\'andez-Garc\'ia}\ \emph
  {et~al.}(2013)\citenamefont {Hern\'andez-Garc\'ia}, \citenamefont {Pic\'on},
  \citenamefont {San~Rom\'an},\ and\ \citenamefont
  {Plaja}}]{Hernandez-GarciaPRL2013}%
  \BibitemOpen
  \bibfield  {author} {\bibinfo {author} {\bibfnamefont {C.}~\bibnamefont
  {Hern\'andez-Garc\'ia}}, \bibinfo {author} {\bibfnamefont {A.}~\bibnamefont
  {Pic\'on}}, \bibinfo {author} {\bibfnamefont {J.}~\bibnamefont
  {San~Rom\'an}}, \ and\ \bibinfo {author} {\bibfnamefont {L.}~\bibnamefont
  {Plaja}},\ }\href {\doibase 10.1103/PhysRevLett.111.083602} {\bibfield
  {journal} {\bibinfo  {journal} {Phys. Rev. Lett.}\ }\textbf {\bibinfo
  {volume} {111}},\ \bibinfo {pages} {083602} (\bibinfo {year}
  {2013})}\BibitemShut {NoStop}%
\bibitem [{\citenamefont {Miranda}\ \emph {et~al.}(2014)\citenamefont
  {Miranda}, \citenamefont {Kotur}, \citenamefont {Rudawski}, \citenamefont
  {Guo}, \citenamefont {Harth}, \citenamefont {L'Huillier},\ and\ \citenamefont
  {Arnold}}]{MirandaOL2014}%
  \BibitemOpen
  \bibfield  {author} {\bibinfo {author} {\bibfnamefont {M.}~\bibnamefont
  {Miranda}}, \bibinfo {author} {\bibfnamefont {M.}~\bibnamefont {Kotur}},
  \bibinfo {author} {\bibfnamefont {P.}~\bibnamefont {Rudawski}}, \bibinfo
  {author} {\bibfnamefont {C.}~\bibnamefont {Guo}}, \bibinfo {author}
  {\bibfnamefont {A.}~\bibnamefont {Harth}}, \bibinfo {author} {\bibfnamefont
  {A.}~\bibnamefont {L'Huillier}}, \ and\ \bibinfo {author} {\bibfnamefont
  {C.~L.}\ \bibnamefont {Arnold}},\ }\href {\doibase 10.1364/OL.39.005142}
  {\bibfield  {journal} {\bibinfo  {journal} {Opt. Lett.}\ }\textbf {\bibinfo
  {volume} {39}},\ \bibinfo {pages} {5142} (\bibinfo {year}
  {2014})}\BibitemShut {NoStop}%
\bibitem [{\citenamefont {Gallet}(2014)}]{Gallet2014}%
  \BibitemOpen
  \bibfield  {author} {\bibinfo {author} {\bibfnamefont {V.}~\bibnamefont
  {Gallet}},\ }\emph {\bibinfo {title} {Dispositifs exp\'erimentaux pour la
  caract\'erisation spatio-temporelle de chaines laser femtosecondes
  haute-puissance}},\ \href@noop {} {Ph.D. thesis},\ \bibinfo  {school} {Orsay
  University} (\bibinfo {year} {2014})\BibitemShut {NoStop}%
\bibitem [{\citenamefont {Miranda}\ \emph {et~al.}(2012)\citenamefont
  {Miranda}, \citenamefont {Fordell}, \citenamefont {Arnold}, \citenamefont
  {L'Huillier},\ and\ \citenamefont {Crespo}}]{MirandaOE2012}%
  \BibitemOpen
  \bibfield  {author} {\bibinfo {author} {\bibfnamefont {M.}~\bibnamefont
  {Miranda}}, \bibinfo {author} {\bibfnamefont {T.}~\bibnamefont {Fordell}},
  \bibinfo {author} {\bibfnamefont {C.}~\bibnamefont {Arnold}}, \bibinfo
  {author} {\bibfnamefont {A.}~\bibnamefont {L'Huillier}}, \ and\ \bibinfo
  {author} {\bibfnamefont {H.}~\bibnamefont {Crespo}},\ }\href {\doibase
  10.1364/OE.20.000688} {\bibfield  {journal} {\bibinfo  {journal} {Opt.
  Express}\ }\textbf {\bibinfo {volume} {20}},\ \bibinfo {pages} {688}
  (\bibinfo {year} {2012})}\BibitemShut {NoStop}%
\bibitem [{\citenamefont {Berry}(2004)}]{BerryJoOAPaAO2004}%
  \BibitemOpen
  \bibfield  {author} {\bibinfo {author} {\bibfnamefont {M.~V.}\ \bibnamefont
  {Berry}},\ }\href {http://stacks.iop.org/1464-4258/6/i=2/a=018} {\bibfield
  {journal} {\bibinfo  {journal} {J. Opt A: Pure Appl. Op.}\ }\textbf {\bibinfo
  {volume} {6}},\ \bibinfo {pages} {259} (\bibinfo {year} {2004})}\BibitemShut
  {NoStop}%
\end{thebibliography}
\end{document}